\documentclass[12pt]{article}
\usepackage{amssymb}

\begin{document}
\begin{center}
{\bf NEW TWO-DIMENSIONAL QUANTUM MODELS PARTIALLY SOLVABLE BY
SUPERSYMMETRICAL APPROACH}\\
\vspace{0.5cm} {\large \bf M.V. Ioffe$^{}$\footnote{E-mail:
m.ioffe@pobox.spbu.ru},
P.A. Valinevich$^{}$\footnote{E-mail: pasha@PV7784.spb.edu}}\\
\vspace{0.2cm}
Department of Theoretical Physics, Sankt-Petersburg State University,\\
198504 Sankt-Petersburg, Russia

\end{center}
\vspace{0.2cm} \hspace*{0.5in} \hspace*{0.5in}
\begin{minipage}{5.0in}
{\small New solutions for second-order intertwining relations in
two-dimensional SUSY QM are found via the repeated use of the first
order supersymmetrical transformations with intermediate constant
unitary rotation. Potentials obtained by this method -
two-dimensional generalized P\"oschl-Teller potentials - appear to be
shape-invariant. The recently proposed
method of $SUSY-$separation of variables is implemented to obtain
a part of their spectra, including the ground state.
Explicit expressions for energy
eigenvalues and corresponding normalizable eigenfunctions are
given in an analytic form. Intertwining relations of higher orders are
discussed.
\\
\vspace*{0.1cm} PACS numbers: 03.65.-w, 03.65.Fd, 11.30.Pb }
\end{minipage}
\vspace*{0.2cm}
\section*{1. Introduction}
\vspace*{0.5cm} \hspace*{3ex} The importance of each new exactly
solvable model in one-dimensional (1D) Quantum Mechanics is well
known, especially because the list of such models is quite small.
The elegant modern approach used for the study and classification of
these ''elite'' models was provided by Supersymmetrical Quantum
Mechanics (SUSY QM) \cite{wit},\cite{cooper}, which is in essence
an alternative formulation of the famous Factorization Method
\cite{infeld} in one-dimensional Quantum Mechanics. Furthermore,
the introduction in the framework of SUSY QM of a new notion - the
shape invariance \cite{shinv},\cite{cooper} - gave a novel,
algebraic, tool to deal with such kind of models. There are
different ways of going beyond the scope of the standard
Witten's SUSY QM in order to enlarge the class of involved
models. The Higher Order SUSY QM (HSUSY QM), or, equivalently,
Polynomial and $N-$fold SUSY QM \cite{acdi},\cite{highod} as well
as constructions for multidimensional coordinate spaces
\cite{2d},\cite{abei} are among the most promising ones.

From the very beginning, after 1D SUSY QM was formulated by
Witten \cite{wit}, the question of finding the opportunity to generalize it for higher
dimensions of space attracted considerable attention. A direct
$d-$dimensional generalization was built in \cite{abei} by means
of methods originating from SUSY Quantum Field Theory. In this
approach the Superhamiltonian (of block-diagonal form) includes
both scalar and matrix components and can be used to analyse
different physical problems with matrix potentials \cite{neelov}.

In the particular case of two-dimensional space an alternative
SUSY QM approach was proposed, which directly generalizes the
HSUSY QM ideas, namely, the use of the SUSY intertwining
relations with the second order supercharges. This method avoids an
appearance of matrix potentials and provides the intertwining of
two scalar Schr\"odinger Hamiltonians. A large class of such
intertwined Hamiltonians was found in \cite{david1} - \cite{david4}.

In the framework of the latter approach two new methods for the
study of the spectra and the (normalizable) eigenfunctions of
the two-dimensional quantum models were proposed recently in
\cite{newmet}, \cite{pseudo}: $SUSY-$separation of variables and
the two-dimensional shape invariance (see also the review-like paper
\cite{ioffe}). The combination of both of them was explored to
investigate a specific model - a generalized 2D Morse potential
with three free parameters - which is not amenable to
the conventional separation of variables. As a result, this model
turned out to be {\bf partially solvable}, i.e. only a part of
the variety of its {\bf normalizable} wave functions and
corresponding values of energies were found {\bf analytically}.
Thus the transfer from one-dimensional to two-dimensional shape
invariance was accompanied by the loss of the complete solvability
with only the partial one remaining. It is worth mentioning here that
each of the 2D Hamiltonians involved in the second order
intertwining relation is integrable: the symmetry operator of
the fourth order in derivatives was constructed explicitly in terms
of supercharges \cite{david1}, \cite{david3}.

In this paper both approaches of two-dimensional SUSY QM - the direct
two-dimensional generalization \cite{2d},\cite{abei} and the second order
construction \cite{david1}-\cite{david4} -
will be used to build and to investigate some new models, to
which no standard separation of variables can be applied. Again
$SUSY-$separation of variables turns out to be applicable to the model, providing a set
of normalizable wave functions. This model, which is shown to be partially
solvable, will be called a {\bf 2D-generalized P\"oschl-Teller potential}.

As for the method of the two-dimensional shape invariance
\cite{newmet} - \cite{ioffe}, the situation is more delicate.
Though the considered model possesses the property of shape invariance,
the corresponding solutions of the Schr\"odinger equation turned out to be
unnormalizable.

The paper is organized as follows. In Section 2 the known methods
of 2D SUSY QM will be described briefly in order to simplify
the comprehension of the new results. A new technique of searching for solutions of
the two-dimensional second order intertwining relations will be
presented in Section 3, and, in particular, two-dimensional
generalizations of P\"oschl-Teller potentials will be constructed.
In Section 4 $SUSY-$separation of variables will be used to find a part of the
spectrum of this model and analytical expressions for its wave
functions, including the ground state. The peculiarities of shape invariance
are also investigated. In Section 5 an additional
structure with two different superpartners for the same
Hamiltonian is presented, new intertwining relations of fourth
and sixth orders in derivatives are constructed (the last ones are
shape-invariant).

\section*{2. Basics of 2D SUSY QM}

\subsection*{2.1. 2D representation of SUSY algebra}

\par The SUSY algebra of quantum mechanics is given by the
following (anti)commutation relations \cite{wit}:
\begin{equation} \{ \hat{Q}^+, \hat{Q}^- \} = \hat{H};\qquad \{
\hat{Q}^+, \hat{Q}^+ \} = \{ \hat{Q}^-, \hat{Q}^- \} = 0;\qquad [
\hat{Q}^\pm, \hat{H} ] = 0.\label{alg}
\end{equation}
In the case of two dimensions it can be realized \cite{2d},
\cite{abei} by the following 4$\times$4 matrix operators:
\begin{equation} \hat H = \left(
\begin{array}{ccc}H^{(0)}(\vec x)&0&0\\ 0& H^{(1)}_{ik}(\vec
x)&0\\0&0&H^{(2)}(\vec x)
\end{array} \right);\quad i,k=1,2;\quad \hat Q^+ = (\hat Q^-)^{\dagger} =
\left(
\begin{array}{cccc}0&0&0&0\\
q_1^-&0&0&0\\q_2^-&0&0&0\\0&p_1^+&p_2^+&0
\end{array} \right).\label{suham}
\end{equation}
where two scalar Hamiltonians $H^{(0)}, H^{(2)}$ and one
2$\times$2 matrix Hamiltonian $H^{(1)}_{ik}$ of Schr\"odinger type
can be expressed in a quasifactorized form (compare to the factorized
form in one-dimensional case \cite{wit}, \cite{cooper}):
\begin{eqnarray}
H^{(0)}&=&q_l^+q_l^- = -\partial_l^2 + V^{(0)}({\vec x}) =
-\partial_l^2 + \Bigl(\partial_l\chi(\vec x)\Bigr)^2 -
\partial_l^2\chi(\vec x);\,\, \partial_l^2 \equiv
\partial_1^2 + \partial_2^2; \nonumber\\
H^{(2)}&=&p_l^+p_l^-=-\partial_l^2 + V^{(2)}({\vec x}) =
-\partial_l^2 + \Bigl(\partial_l\chi(\vec x)\Bigr)^2 +
\partial_l^2\chi(\vec x);\label{hams}\\
H^{(1)}_{ik}&=&q_i^-q_k^+ + p_i^-p_k^+ = -\delta_{ik}
\partial_l^2 + \delta_{ik}\Bigl(\bigl(\partial_l\chi(\vec x)\bigr)^2 -
\partial_l^2\chi(\vec x)\Bigr) + 2\partial_i\partial_k
\chi(\vec x),\nonumber
\end{eqnarray}
with components of supercharges of first order in derivatives:
\begin{equation}
q_l^{\pm}\equiv \mp\partial_l + \partial_l \chi(\vec x);\quad
p_l^{\pm}\equiv \epsilon_{lk}q_k^{\mp},\label{2p}
\end{equation}
where $\partial_i \equiv \partial/\partial x_i$ and summation over
repeated indices is implied.
Anticommutators in (\ref{alg}) produce the following intertwining
relations for the component Hamiltonians $H^{(0)}, H^{(1)}_{ik},
H^{(2)}$ of the Superhamiltonian (\ref{suham}):
\begin{equation}
H^{(0)}q_i^+=q_k^+H_{ki}^{(1)};\quad H_{ik}^{(1)}q_k^- =
q_i^-H^{(0)}; \quad H^{(1)}_{ik}p_k^-=p_i^-H^{(2)} ; \quad
H^{(2)}p_i^+ = p_k^+H^{(1)}_{ki}.\label{2dintw}
\end{equation}
They connect the spectrum of the matrix hamiltonian with spectra
of two scalar ones. In general, $H^{(0)}$ and $H^{(2)}$ {\it are
not} isospectral since $q^+_kp^-_k \equiv 0$ due to (\ref{2p}).

\subsection*{2.2. Second-order supercharges in 2D SUSY QM}

\par Two-dimensional SUSY QM models without any matrix Hamiltonians
were constructed \cite{david1}, \cite{david2}, \cite{ioffe} by means of second
order supercharges
\begin{equation}
Q^+ = (Q^-)^\dag = g_{ik}(\vec{x})\partial_i\partial_k +C_i
(\vec{x})\partial_i + B(\vec{x}) \label{Q}
\end{equation}
where $g_{ik}, C_i, B$ are arbitrary real functions. Some
particular solutions for two scalar Hamiltonians $H^{(0),(1)}$
which satisfy the intertwining relations
\begin{equation}
H^{(1)}(\vec{x})Q^+ = Q^+H^{(0)}(\vec{x});\qquad
H^{(0)}(\vec{x})Q^- = Q^-H^{(1)}(\vec{x}), \label{intw2d}
\end{equation}
were found. They both possess the symmetry operators $R^{(1,2)}$
of fourth order in derivatives \cite{david1}, \cite{david3}:
$$
[R^{(i)}, H^{(i)}] = 0;\quad i = 0,1;\qquad R^{(0)} = Q^-Q^+;
\qquad R^{(1)} = Q^+Q^-,
$$
which are not, in general, polynomials of $H^{(i)}$.
\par In terms of unknown functions $g_{ik}, C_i, B, V^{(0)}, V^{(1)}$
Eq.(\ref{intw2d}) has the form \cite{david1} of seven nonlinear
partial  differential equations, and its general solution is not
known. To obtain particular solutions different ans\"atze for
''metrics'' $g_{ik}$ were used.
\par Only the choice of Laplacian  (elliptic) metrics $g_{ik}
(\vec{x}) = diag(1, 1)$ leads to Hamiltonians amenable to
$R-$separation \cite{miller} of variables. All other choices of
metrics give nontrivial results. The case  of Lorentz
(hyperbolic) metrics $g_{ik} = diag(1,-1)$ was investigated in
papers \cite{david1} - \cite{david4}. In particular, the intertwining relations
(\ref{intw2d}) were reduced to the pair of differential equations:
\begin{equation}
\partial_1\partial_2 F = 0;\qquad\partial_-(C_-F)  = -
\partial_+(C_+F).\label{2odmain}
\end{equation}
where\footnote{We use here the definition of $x_\pm$ slightly
different from the analogous one in \cite{david2} - \cite{david4}.} $x_\pm = (x_1 \pm
x_2)/\sqrt{2}$ and $C_{1,2}$ were proven to satisfy $C_\pm \equiv C_1
\mp C_2 \equiv C_\pm(\sqrt{2}x_\pm)$. Then potentials
$V^{(0),(1)}$ and the supercharges $Q^+$  are expressed in terms
of functions $C_\pm(\sqrt{2}x_\pm)$ and $F(\vec{x})$ which obviously can
be written as $F= F_1(2x_1)+ F_2(2x_2)$ according to (\ref{2odmain}):
\begin{eqnarray}
V^{(0),(1)} & = & \mp \frac{1}{2}\Bigl(C'_+(\sqrt{2}x_+) +
C'_-(\sqrt{2}x_-)\Bigr) + \frac{1}{8}\Bigl(C_+^2(\sqrt{2}x_+) +
C_-^2(\sqrt{2}x_-)\Bigr)  +\nonumber\\
&&+\frac{1}{4}\Bigl(F_2(2x_2) - F_1(2x_1)\Bigr); \label{ve2od}\\
Q^+ & = & (\partial_1^2 - \partial_2^2) + C_1\partial_1 +
C_2\partial_2 + B;\label{quplus}\\ B & = &
\frac{1}{4}\Bigl(C_+(\sqrt{2}x_+)C_-(\sqrt{2}x_-) + F_1(2x_1) +
F_2(2x_2)\Bigr),\label{be2od}
\end{eqnarray}
where the prime denotes the derivative of function with respect to its
argument. A list of particular solutions of (\ref{2odmain}) was
obtained in \cite{david2} - \cite{david4}. In the next Section we will obtain new
solutions for the case of hyperbolic metrics.

\section*{3. New solutions for the Lorentz (hyperbolic) metrics}

\subsection*{3.1. Intertwining of second order with reducible supercharges}

Let us consider two Superhamiltonians $\hat{H}$ and
$\hat{\widetilde{H}}$ of 2D SUSY QM:
\begin{equation}
\begin{array}{ccc}
\hat H = \left(
\begin{array}{ccc}H^{(0)}(\vec x)&0&0\\ 0& H^{(1)}_{ik}(\vec
x)&0\\0&0&H^{(2)}(\vec x);
\end{array} \right); & \quad &
\hat{\widetilde{H}} = \left(
\begin{array}{ccc}\widetilde{H}^{(0)}(\vec x)&0&0\\ 0&
\widetilde{H}^{(1)}_{ik}
(\vec x)&0\\0&0&\widetilde{H}^{(2)}(\vec x);
\end{array} \right),
\end{array}
\end{equation}
with superpotentials $\chi(\vec{x})$ and
$\widetilde{\chi}(\vec{x})$, correspondingly.
\par In addition, let $H^{(1)}_{ik}$ and $\widetilde{H}^{(1)}_{ik}$ be
linked by an unitary $2\times2$ matrix transformation $U$:
\begin{equation}
U_{ik}^{\phantom{x}}\widetilde{H}^{(1)}_{kl} =
H^{(1)}_{im}U_{ml}^{\phantom{x}};
\end{equation}
\begin{equation}
U = \alpha_0\sigma_0 +
i\overrightarrow{\alpha}\overrightarrow{\sigma}; \quad \alpha_0^2
+ \overrightarrow{\alpha}^2 = 1; \quad \alpha_0, \alpha_i \in
\bf{R},\label{om}
\end{equation}
where $\sigma_i$ are the Pauli matrices and $\sigma_0$ is the unit
matrix.
\par Then (due to (\ref{2dintw})) the scalar Hamiltonians $H^{(0)}$ and
$\widetilde{H}^{(0)}$ can be included in the chain:
\begin{equation}
H^{(0)} \stackrel{q_l^{\pm}}{\leftarrow\rightarrow}
H^{(1)}_{ik}\stackrel{U_{lm}}{\leftarrow\rightarrow}
\widetilde{H}^{(1)}_{ik}\stackrel{\tilde{q}_m^{\mp}}
{\leftarrow\rightarrow} \widetilde{H}^{(0)}, \label{chainh}
\end{equation}
leading to the intertwining relations between a pair of
scalar Hamiltonians:
\begin{equation}
H^{(0)}Q^- = Q^-\widetilde{H}^{(0)},\qquad Q^+H^{(0)} =
\widetilde{H}^{(0)}Q^+ \label{intertw}
\end{equation}
with second order operators
\begin{equation}
Q^- = (Q^+)^\dag = q^+_iU_{ik}\tilde{q}^-_k.\label{ku}
\end{equation}
This intertwining operator is constructed from two first order
ones with intermediate matrix transformation $U_{ik}$. Precisely
this matrix provides that such
supercharges $Q^\pm$ are nontrivial and, contrary to Subsection 2.1., can be
naturally described as {\bf reducible} (compare with the case of
one-dimensional reducibility introduced in \cite{acdi}).
\par In contrast to the approach of \cite{newmet} (see Subsection
2.2.), the first Hamiltonian in the chain (\ref{chainh}) is
quasifactorized according to Eq.(\ref{hams}). Therefore the
solution of the corresponding Schr\"odinger equation with zero
energy can be written as $\Psi_0^{(0)}\sim \exp{(-\chi)}$. Due to
expression (\ref{ku}), $\exp{(-\chi)}$ is a zero mode of
supercharge $Q^+$ as well. In general, until the specific form of
$\chi(\vec{x})$ is chosen, the normalizability of this solution
is not guaranteed. But in the concrete model (\ref{pot2new})
analyzed below in Subsection 3.3. the zero energy solution is
{\it normalizable} due to asymptotic properties of
$\chi{(\vec{x})}$ for corresponding ranges (\ref{ranpar}) of
parameters.
\par Target Hamiltonians $H^{(0)}$ and $\widetilde{H}^{(0)}$
are expressed in terms of two unknown functions $\chi$ and
$\tilde{\chi}$ (see (\ref{hams})). To determine these functions
one should substitute (\ref{hams}), (\ref{om}) and (\ref{ku})
into (\ref{intertw}). After some manipulations one obtains the
system of equations for $\chi_\pm = (\chi \pm \tilde{\chi})/2$:
\begin{eqnarray}
\alpha_3\Box\chi_- +2\alpha_1\partial_1\partial_2\chi_- = 0;
\qquad \alpha_1\Box\chi_+ - 2\alpha_3\partial_1\partial_2\chi_+
&= 0;\label{lin1}\\
\alpha_2\Box\chi_+ + 2\alpha_0\partial_1\partial_2\chi_- = 0;
\qquad \alpha_0\Box\chi_- + 2\alpha_2\partial_1\partial_2\chi_+
&= 0; \label{lin2}
\end{eqnarray}
\begin{equation}
(\partial_k \chi_-)(\partial_k\chi_+) = 0, \label{nonlin}
\end{equation}
where $\Box \equiv \partial_1^2 - \partial_2^2$. Eq.
(\ref{nonlin}) (which is equivalent to $(\partial_k\chi)^2 =
(\partial_k\widetilde{\chi})^2$) can be used to simplify expressions
(\ref{hams}) for the Hamiltonians $H^{(0)}$ and
$\widetilde{H}^{(0)}$:
\begin{equation}
H^{(0)}, \widetilde{H}^{(0)}  = -\partial_l^2 + \Bigl(
(\partial_l\chi_+)^2 -
\partial_l^2\chi_+ \Bigr) + \Bigl( (\partial_l\chi_-)^2 \mp
\partial_l^2\chi_- \Bigr).\label{potchi01}
\end{equation}
\par Linear partial differential equations (\ref{lin1})-(\ref{lin2})
can be easily solved, but the solution of the nonlinear Eq.
(\ref{nonlin}) is a nontrivial problem.

\subsection*{3.2. The particular solutions of the intertwining
relations}

\par It can be shown that in the case when {\bf all coefficients}
$\alpha_i$ and $\alpha_0$ in (\ref{om}) do not vanish, potentials
$V^{(0)}$ and $\widetilde{V}^{(0)}$ are 4-th order polynomials on
$x_{1,2}$ with some additional constraints for their
coefficients. In the present paper we will consider potentials
beyond this rather narrow class, restricting ourselves to the
{\bf particular} case $\alpha_0 = \alpha_1 = \alpha_2 = 0;\quad
\alpha_3 \ne 0$, i.e. $U = \sigma_3$. Then the metrics of
supercharges $Q^\pm$ is Lorentz, i.e. they belong to the class
discussed in Subsection 2.2.
\par For this case equations (\ref{lin1}) - (\ref{lin2}) read:
$$
\Box\chi_-  =  0;\qquad
\partial_1\partial_2\chi_+  =  0.
$$
Their solution is
\begin{eqnarray}
\chi_- & = & \mu_+(x_+) + \mu_-(x_-),\nonumber\\
\chi_+ & = & \mu_1(x_1) + \mu_2(x_2), \nonumber
\end{eqnarray}
with $\mu_{1,2}$, $\mu_\pm$ being arbitrary functions. The last
equation (\ref{nonlin}) takes the form
$$
\mu'_1(x_1)\left[ \mu'_+(x_+) + \mu'_-(x_-) \right] + \mu'_2(x_2)
\left[ \mu'_+(x_+) - \mu'_-(x_-) \right] = 0.
$$
By substitutions $\phi \equiv \mu'$, it becomes purely functional
(without derivatives) equation:
\begin{equation}
\phi_1(x_1)\left[ \phi_+(x_+) + \phi_-(x_-) \right] = -
\phi_2(x_2) \left[ \phi_+(x_+) - \phi_-(x_-) \right].
\label{phimain}
\end{equation}
The {\bf general} solution of (\ref{phimain}) is given in the
Appendix\footnote{It was derived by D.N. Nishnianidze (private
communication).}. Some particular cases will be discussed in
Subsection 3.3.
\par The Hamiltonians (\ref{potchi01}) and intertwining
operators (\ref{ku}) can be expressed in terms of $\phi$:
\begin{eqnarray}
V^{(0)}, \widetilde{V}^{(0)} =& \Bigl(\phi_1^2(x_1) -
\phi'_1(x_1)\Bigr) + \Bigl(\phi_2^2(x_2) - \phi'_2(x_2)\Bigr) +
\Bigl(\phi_+^2(x_+) \mp
\phi'_+(x_+)\Bigr) +\nonumber\\
&+\Bigl(\phi_-^2(x_-) \mp \phi'_-(x_-)\Bigr),\label{potphiold}\\
Q^\pm = &\partial_1^2 - \partial_2^2 \pm \sqrt{2}\Bigl(\phi_+(x_+)
+ \phi_-(x_-)\Bigr)\partial_1 \mp\sqrt{2} \Bigl(\phi_+(x_+) -
\phi_-(x_-)\Bigr)\partial_2 - \nonumber\\
& -\Bigl(\phi_1^2(x_1) - \phi'_1(x_1)\Bigr) + \Bigl(\phi_2^2(x_2)
- \phi'_2(x_2)\Bigr) +2\phi_+(x_+)\phi_-(x_-).\nonumber
\end{eqnarray}
\par By rearrangement of terms Eq.(\ref{phimain}) can be rewritten as:
\begin{equation}
\phi_+(x_+)[\phi_1(x_1) + \phi_2(x_2)] = -\phi_-(x_-)[\phi_1(x_1)
- \phi_2(x_2)],\label{1}
\end{equation}
i.e. in the form similar to the initial Eq.(\ref{phimain}). This
means that (\ref{phimain}) possesses the symmetry property (which
will be called $S_1$ symmetry in the subsequent text): if
$\biggl\{ \phi_1(x_1), \phi_2(x_2), \phi_+(x_+), \phi_-(x_-) \biggr\}$ is a
solution, then $\biggl\{\phi_+(x_1), \phi_-(x_2), \phi_1(x_+),
\phi_2(x_-)\biggr\}$ is also a solution. Let us mention one more discrete
symmetry of (\ref{nonlin}), $S_2$ symmetry: $\biggl\{\phi_1(x_1),
\phi_2(x_2), \phi_+(x_+), \phi_-(x_-)\biggr\} \longrightarrow
\biggl\{\phi_1(x_1), -\phi_2(x_2), \phi_+^{-1}(x_+), \phi_-^{-1}(x_-)\biggr\}$.
The $S_1$-symmetry produces another supersymmetrical model:
\begin{eqnarray}
\mathcal{V}^{(0)}, \widetilde{\mathcal{V}}^{(0)} =&
\Bigl(\phi_1^2(x_+) \mp \phi'_1(x_+)\Bigr) + \Bigl(\phi_2^2(x_-)
\mp \phi'_2(x_-)\Bigr) + \Bigl(\phi_+^2(x_1) - \phi'_+(x_1)\Bigr)
+\nonumber\\ &+\Bigl(\phi_-^2(x_2)
- \phi'_-(x_2)\Bigr),\label{potphinew}\\
\mathcal{Q}^\pm = &\partial_1^2 -
\partial_2^2 \pm \sqrt{2}\Bigl(\phi_1(x_+) +
\phi_2(x_-)\Bigr)\partial_1 \mp\sqrt{2} \Bigl(\phi_1(x_+) -
\phi_2(x_-)\Bigr)\partial_2 - \nonumber\\
& -\Bigl(\phi_+^2(x_1) - \phi'_+(x_1)\Bigr) + \Bigl(\phi_-^2(x_2)
- \phi'_-(x_2)\Bigr) + 2\phi_1(x_+)\phi_2(x_-).\nonumber
\end{eqnarray}
Below both forms (\ref{potphiold}) and (\ref{potphinew}) will be
explored.
\par To compare the new notations of this  Section with those
of \cite{david2} - \cite{david4} (see Subsection 2.2.) one can use the following
relations:
\begin{equation}
\begin{array}{rcl}
C_\pm(\sqrt{2}x_\pm) & = & 2\sqrt{2}\phi_\pm(x_\pm);\\
F_{1,2}(2x_{1,2}) & =  & \mp 4\Bigl( \phi_{1,2}^2(x_{1,2}) -
\phi'_{1,2}(x_{1,2}) \Bigr).
\end{array} \nonumber
\end{equation}

\subsection*{3.3. Nonperiodical solutions for potentials $V^{(0)}$,
$\widetilde{V}^{(0)}$.}

\par From Eq.(\ref{ell}) one can conclude that for an arbitrary
choice of the parameters $a, b, c$ the functions $\phi_{1,2}$ are expressed in
terms of elliptic (Jacobi or Weierstrass) functions \cite{bather}
(volume 3). In this paper we restrict ourselves by considering
the limiting cases, for which the potentials are not periodical
(the models with periodicity properties in $x_{1,2}$ will be
studied elsewhere).
\par The integral in the r.h.s. in (\ref{ell}) is an elementary function only
if either some of coefficients $a, b, c$ are zero or the quadratic
polynomial is a full square. There are two
families of solutions of (\ref{phimain}), with members
interconnected by symmetries $S_1$ and $S_2$ (and their
combinations):
$$ {\bf(i)}\qquad\qquad\qquad \phi_1(x) =
\phi_2(x) = A/x;\,\,\phi_+ = \phi_- = B/x\quad (A, B = const).$$
The multiparticle potentials of this type were found
in \cite{tanaka} to be quasi-exactly solvable \cite{turb}.
All other members of this family allow the separation of
variables.
\begin{eqnarray}
{\bf (ii)} \qquad\qquad &\phi_1(x) &= \phi_2(x)  =  M \left(
\delta_+e^{\alpha x} + \delta_-e^{-\alpha x}\right);\nonumber\\
&\phi_+(x) & =  -L\frac{\delta_+e^{\alpha x/\sqrt{2}} -
\delta_-e^{-\alpha x/\sqrt{2}}}{\delta_+e^{\alpha x/\sqrt{2}} +
\delta_-e^{-\alpha x/\sqrt{2}}};\quad \phi_-(x)  =  L\coth{\left(
\alpha x/\sqrt{2} \right)}.\label{genmorse}
\end{eqnarray}
For the particular case $\delta_-=0$ one has:
\begin{eqnarray}
V^{(0)}, \widetilde{V}^{(0)} &= (B^2e^{-2\alpha x_1} + B\alpha
e^{-\alpha x_1}) + (B^2e^{-2\alpha x_2} + B\alpha e^{-\alpha
x_2}) + \nonumber\\ & 4A^2 + A(2A \mp
\alpha)\left[\sinh{\left(\frac{\alpha}{2}(x_1 -
x_2)\right)}\right]^{-2} \label{morse}
\end{eqnarray}
with two new constants $A, B$ instead of $M, L, \delta_+$. This
potentials (up to translations in $x_{1,2}$) were analyzed in
\cite{newmet} and were found to be shape-invariant.
\par Another particular case $\delta_+=-\delta_-$ for (\ref{genmorse})
after using symmetries and redefinition of parameters gives:
\begin{eqnarray}
\phi_1 &= -\phi_2 = &\frac{A}{\sinh{\sqrt{2}\alpha x}}\nonumber\\
\phi_+ &= \phi_- = &B \tanh{\alpha x}, \label{phisol}
\end{eqnarray}
The corresponding potentials and intertwining operators for this
model due to (\ref{potphiold}) are:
\begin{eqnarray}
V^{(0)}, \widetilde{V}^{(0)} &=& \left( B^2 - \frac{B(B \pm
\alpha)}{\cosh^2{(\frac{\alpha}{\sqrt{2}}(x_1 + x_2))}} \right) +
\left( B^2 - \frac{B(B \pm
\alpha)}{\cosh^2{(\frac{\alpha}{\sqrt{2}}
(x_1 - x_2))}} \right) +\nonumber\\
&&+A\left[ \frac{A - \sqrt{2}\alpha\cosh{(\sqrt{2}\alpha
x_1)}}{\sinh^2{(\sqrt{2}\alpha x_1)}} + \frac{A + \sqrt{2}\alpha
\cosh{(\sqrt{2}\alpha x_2)}}{\sinh^2{(\sqrt{2}\alpha
x_2)}}\right], \label{pot1}\\
Q^\pm &=& \partial_1^2 - \partial_2^2 \pm\sqrt{2}B\Biggl[
\tanh{\left(\frac{\alpha}{\sqrt{2}}(x_1 + x_2)\right)} +
\tanh{\left(\frac{\alpha}{\sqrt{2}}(x_1 - x_2)\right)}
\Biggr]\partial_1
\mp\nonumber\\
&&\mp\sqrt{2}B\Biggl[ \tanh{\left(\frac{\alpha}{\sqrt{2}}(x_1 +
x_2)\right)} - \tanh{\left(\frac{\alpha}{\sqrt{2}}(x_1 -
x_2)\right)} \Biggr]\partial_2 - \nonumber\\ &&- A\left[ \frac{A -
\sqrt{2}\alpha\cosh{(\sqrt{2}\alpha
x_1)}}{\sinh^2{(\sqrt{2}\alpha x_1)}} - \frac{A + \sqrt{2}\alpha
\cosh{(\sqrt{2}\alpha x_2)}}{\sinh^2{(\sqrt{2}\alpha
x_2)}}\right] +\nonumber\\
&&+ 2B^2\tanh{\left(\frac{\alpha}{\sqrt{2}}(x_1 +
x_2)\right)}\tanh{\left(\frac{\alpha}{\sqrt{2}}(x_1 -
x_2)\right)}.\nonumber
\end{eqnarray}
Solution (\ref{potphinew}), obtained by the discrete symmetry $S_1$,
is:
\begin{eqnarray}
\mathcal{V}^{(0)}, \widetilde{\mathcal{V}}^{(0)} &=& \left( B^2 -
\frac{B(B + \alpha)}{\cosh^2{(\alpha x_1)}} \right) + \left( B^2
- \frac{B(B + \alpha)}{\cosh^2{(\alpha x_2)}}
\right) +\nonumber\\
&&+A\left[ \frac{A \mp \sqrt{2}\alpha\cosh{(\alpha(x_1 +
x_2))}}{\sinh^2{(\alpha(x_1 + x_2))}} + \frac{A \pm
\sqrt{2}\alpha\cosh{(\alpha(x_1 - x_2))}}{\sinh^2{(\alpha(x_1 -
x_2))}}\right]\label{pot2new}\\
\mathcal{Q}^\pm &=& \partial_1^2 - \partial_2^2 \pm\sqrt{2}A\left[
\frac{1}{\sinh{(\alpha(x_1 + x_2))}} + \frac{1}{\sinh{(\alpha(x_1
- x_2))}} \right]\partial_1 \mp \nonumber\\
&&\mp\sqrt{2}A\left[ \frac{1}{\sinh{(\alpha(x_1 + x_2))}} -
\frac{1}{\sinh{(\alpha(x_1 - x_2))}} \right]\partial_2-\nonumber\\
&&- \left[ B^2 - \frac{B(B + \alpha)}{\cosh^2{(\alpha x_1)}}
\right] + \left[B^2 - \frac{B(B + \alpha)}{\cosh^2{(\alpha
x_2)}}\right]+\nonumber\\
&&+\frac{2A^2}{\sinh{(\alpha(x_1 + x_2))}\sinh{(\alpha(x_1 -
x_2))}}\nonumber
\end{eqnarray}
\par Both potentials (\ref{pot1}) and (\ref{pot2new}) can be
treated as superpositions of two one-dimensional P\"oschl-Teller
terms plus a singular term (so we will refer to them as {\bf
2D-generalized P\"oschl-Teller potentials}). Each of them
possesses a term which prevents application of the conventional method
of separation of variables to determine their eigenfunctions and
eigenvalues. Meanwhile, a part of the spectrum and corresponding
eigenfunctions will be found by the method of
{\it $SUSY-$separation of variables} (see \cite{newmet}, \cite{ioffe}) in the next
Section\footnote{Other members of the same family ($S_2$- and
$S_2S_1$-symmetric to (\ref{phisol})) can be treated
analogously.}.

\section*{4. Partial solvability of 2D-generalized
P\"oschl-Teller potentials}

\subsection*{4.1. $SUSY-$separation of variables}

As far as Hamiltonians with potentials (\ref{pot2new}) are
intertwined by operators $\mathcal{Q}^{\pm}$ with Lorentz metrics,
we shall briefly remind the reader of the general method for
searching for eigenvalues and eigenfunctions proposed in
\cite{newmet}. From intertwining relations
$\mathcal{Q}^+\mathcal{H}^{(0)} = \widetilde{\mathcal{H}}^{(0)}
\mathcal{Q}^+$ (where $\mathcal{H}^{(0)} = -\partial_l^2 +
\mathcal{V}^{(0)}$ and $\widetilde{\mathcal{H}}^{(0)} =
-\partial_l^2 + \widetilde{\mathcal{V}}^{(0)}$) one obtains that
the subspace of zero modes\footnote{Here we suppose that $(N+1)$
normalizable zero modes $\Omega_n(\vec x)$ are known, and
$\vec{\Omega}(\vec{x})$ is a column vector with components
$\Omega_n(\vec{x})$, $n = 0, 1, \ldots N$.} of the supercharge
$\mathcal{Q}^+$:
\begin{equation}
\mathcal{Q}^+\vec{\Omega}(\vec{x}) = 0,\label{nom}
\end{equation}
is closed under the
action of $\mathcal{H}^{(0)}$:
\begin{equation}
\mathcal{H}^{(0)}\vec{\Omega}(\vec{x}) =
\hat{C}\vec{\Omega}(\vec{x}) \label{clos}
\end{equation}
with some constant matrix $\hat C.$
\par To determine the eigenvalues $E_k$ and eigenfunctions
$\Psi_k(\vec x)$ of $\mathcal{H}^{(0)}$ one needs (see more
details in \cite{newmet}) a matrix $\hat B,$ which satisfies the
matrix equation $\hat{B}\hat{C} = \hat{\Lambda}\hat{B}$ with an
unknown yet diagonal matrix $\hat{\Lambda} = diag(\lambda_0,
\lambda_1,\ldots, \lambda_N).$ Then the matrix $\hat{B}$
transforms zero-modes $\Omega_n$'s into wave functions $\Psi_n$'s.
\par Operator $\mathcal{Q}^+$ belongs to type (\ref{quplus}). For
this type of supercharges the problem (\ref{nom}) permits the
conventional separation of variables in $\mathcal{Q}^+$ by means
of "gauge" transformation, which separates variables in the
supercharge:
\begin{equation}
\mathfrak{q}^+ =
e^{-\kappa(\vec{x})}\mathcal{Q}^+e^{\kappa(\vec{x})} =
\partial_1^2 - \partial_2^2 + \frac{1}{4}(F_1(2x_1) + F_2(2x_2)),
\label{kusep}
\end{equation}
\begin{equation}
h(\vec{x}) \equiv e^{-\kappa(\vec{x})}
\mathcal{H}^{(0)}(\vec{x})e^{\kappa(\vec{x})} = -\partial_1^2 -
\partial_2^2 + C_1(\vec{x})\partial_1 - C_2(\vec{x})\partial_2 -
\frac{1}{4}F_1(2x_1) + \frac{1}{4}F_2(2x_2) \label{Hh}.
\end{equation}
$$
\kappa(\vec{x}) \equiv -\frac{\sqrt{2}}{4}\left[ \int
C_+(\sqrt{2}x_+)\,dx_+ + \int C_-(\sqrt{2}x_-)\,dx_- \right].
$$
$$
\omega_n(\vec{x}) = e^{-\kappa(\vec{x})}\,\Omega_n(\vec{x}).
$$
Then the zero modes $\omega_n(\vec{x})$ of $\mathfrak{q}^+$ can
be written as products $\omega_n(\vec{x}) =
\eta_n(x_1)\rho_n(x_2)$, where $\rho_n$ and $\eta_n$ are
eigenfunctions of the one-dimensional Schr\"odinger equations with
"potentials" $(\mp\frac{1}{4}F_{1,2}(2x_{1,2})),$ correspondingly
(see (\ref{kusep})), and common eigenvalues (constants of
separation) $\epsilon_n.$
\par It is obvious that $h\vec{\omega} = \hat{C}\vec{\omega}$ with
the same matrix $\hat{C}$ as in (\ref{clos}). The simplest way to
find $\hat{C}$ is to calculate the r.h.s. of (\ref{Hh}), which
can be rewritten as:
\begin{equation}
h\omega_n = [2\epsilon_n + C_1(\vec{x})\partial_1 -
C_2(\vec{x})\partial_2]\omega_n. \label{homegan}
\end{equation}
As a result, after construction of the matrix $\hat B$ one will
obtain part of the spectrum ${E_k}$ and corresponding wave
functions $\Psi_k(\vec x).$

\subsubsection*{4.1.1. Calculation of $\hat{C}$}

\par This general method, proposed in \cite{newmet} and
used there successfully to investigate the 2D Morse potential, can
be applied to the pair $\mathcal{V}^{(0)}$,
$\widetilde{\mathcal{V}}^{(0)}$ as well. In this case both
one-dimensional equations for multipliers $\eta_n(x_1)$ and
$\rho_n(x_2)$ have the same "potentials" - {\it one dimensional}
P\"oschl-Teller potentials - being exactly-solvable:
\begin{equation}
(-\partial_1^2 + B^2 -B(B + \alpha)\cosh^{-2} {(\alpha
x_1)})\eta_n(x_1) =
\epsilon_n\eta_n(x_1) \label{eta}
\end{equation}
and a similar equation for $\rho_n (x_2).$ By the change of variable
$\xi \equiv \tanh{\alpha x_1}$ Eq.(\ref{eta}) can be reduced to
the generalized Legendre equation \cite{dau} :
$$
\frac{d}{d\xi}\left[ (1-\xi^2)\frac{d\eta}{d\xi} \right] + \left[
s(s+1) - (s^2-\frac{\epsilon}{\alpha^2})\frac{1}{1-\xi^2}
\right]\eta = 0,
$$
where $s = B/\alpha$. To have the finite solution at $\xi = -1$
the condition $\sqrt{(B^2 - \epsilon)/\alpha^2} - s = -n$; $n \in$
{\bf N} must be satisfied. It gives a discrete set of values for
the separation constant $\epsilon$:
$$
\epsilon_n = \alpha^2n(2s - n)
$$
for $n < s$. Corresponding functions (up to normalization
factors) are $\eta_n = P^{s-n}_s(\xi)$, where $P^\mu_\nu(x)$ are
the (generalized) Legendre functions. Thus, one achieves the
expression for $\omega_n$:
$$
\omega_n = P^{s-n}_s(\xi_1)P^{s-n}_s(\xi_2),
$$
with $\xi_i = \tanh{\alpha x_i}$, $i = 1,2$.
\par The next step in calculating the eigenfunctions is evaluating
the r.h.s. of (\ref{homegan}):
\begin{equation}
h\omega_n(\vec{x}) = 2\alpha^2n(2s - n)\omega_n  -
2\sqrt{2}A\alpha(2s-n)(n+1)\frac{(1-\xi_1^2)^{1/2}(1 -
\xi_2^2)^{1/2}}{(\xi_1^2 - \xi_2^2)}\Pi(n, s; \xi_1,
\xi_2),\label{homegan1}
\end{equation}
where me made the shorthand notation
$$
\Pi(n, s; \xi_1, \xi_2) = \xi_2(\xi_1^2 -
1)^{1/2}P^{s-n-1}_s(\xi_1)P^{s-n}_s(\xi_2) - \xi_1(\xi_2^2 -
1)^{1/2}P^{s-n}_s(\xi_1)P^{s-n-1}_s(\xi_2).
$$
Our goal is to represent the r.h.s. of (\ref{homegan1}) as a
linear combination of $\omega_k$'s. For this purpose we use the
reccurrent formula for Legendre functions (see \cite{bather}, v.1,
p.161, eq. (1))
$$
(z^2 - 1)^{1/2}P^{\mu+2}_\nu(z) + 2(\mu + 1)zP^{\mu + 1}_nu(z) =
(z^2 - 1)^{1/2}(\nu - \mu)(\mu + \nu + 1)P^\mu_\nu.
$$
Applying it twice to $\Pi(n, s; \xi_1; \xi_2)$ one obtains the
following reccurrent formula:
\begin{equation}
\Pi(n, s; \xi_1, \xi_2) =
\frac{1}{(n+1)n(2s-n)(2s-n+1)}\Biggl[\frac{2(s-n+1)(\xi_1^2 -
\xi_2^2)}{(1 - \xi_1^2)^{1/2}(1 - \xi_2^2)^{1/2}}\omega_{n-1}+\Pi(n-2,
s; \xi_1, \xi_2)\Biggr]. \label{rekurr1}
\end{equation}
To stop this procedure at $n=0$, one has to consider $s \in
\mathbb{N}$. In this case the Legendre functions turns into the
associate Legendre polynomials, for which $P_n^m(z) \equiv 0$ for
$m>n$. So, applying (\ref{rekurr1}) several times,
$$
\Pi(n, s; \xi_1, \xi_2) = \frac{\xi_1^2 - \xi_2^2}{(1 -
\xi_1^2)^{1/2}(1 - \xi_2^2)^{1/2}}\sum_{k = 0}^{n}a_{nk}\omega_k
$$
with constants $a_{nk}$. The matrix elements $c_{nk}$ of matrix
$\hat{C}$ are:
\begin{equation}
c_{nk} = 2\alpha^2n(2s-n)\delta_{nk} -
2\sqrt{2}A\alpha(2s-n)(n+1)a_{nk};\label{matc}
\end{equation}
\begin{equation}
a_{nk} = \left\{
\begin{array}{ll}
0, & k \ge n;\\
0, &k = n - 2m - 2;\, m = 0,1,2,...\\
2(s-k)\frac{(k-1)!(2s-n-1)!}{(n+1)!(2s-k)!}, & k = n - 2m - 1;\, m
= 0,1,2,...\
\end{array}
\right. \label{matrc}
\end{equation}

\subsubsection*{4.1.2. Calculation of eigenfunctions}

\par Matrix $\hat{C}$  for the model (\ref{pot2new}) appeared
to be triangular, and hence its eigenvalues coincide with the
diagonal elements:
\begin{equation}
E_k = c_{kk} = 2\alpha^2k(2s-k). \label{eigenval}
\end{equation}
This formula gives $E_0 = 0$, demonstrating that the zero energy
solution of $\mathcal{H}^{(0)}$ is a zero mode of $\mathcal{Q}^+$
as well: $\Psi_0 \sim \exp{(-\chi(\vec{x}))}$.
\par In order to avoid zeros on the diagonal of $\hat{C}$ one can shift
Hamiltonians by a constant $\gamma$. This transformation does not
destroy the intertwining relations (\ref{intertw}) and changes
$\hat{C}$ as follows: $c_{ik} \rightarrow c_{ik} +
\gamma\delta_{ik}.$ This new $\hat{C}$ can be diagonalized by the
method, presented in \cite{newmet}. Namely, the formal solution
for the matrix elements of $\hat{B}$ reads:
\begin{equation}
b_{m,p} = b_{m, N-m}\left[ \sum\limits_{l=1}^{N-p-1}(\tau^{(m)})^l
\right]_{N-m,p}\label{formal}
\end{equation}
where $(N+1)$ matrices $\tau^{(m)}$ are defined by
$$
\tau_{n,k}^{(m)} \equiv \frac{c_{n,k}}{c_{N-m, N-m} - c_{k,k}},
$$
and label $(m)$ has values $m = 0, 1, ..., N$. In Eq.(\ref{formal})
the repeated index $N-m$ is {\bf not} summed over,
and (to avoid misunderstanding) $\tau^{{(m)}^l}$ means the $l$th
power of matrix $\tau^{(m)}$. Thus one obtains the recipe for the
construction of eigenfunctions for $\mathcal{H}^{(0)}$ in
(\ref{pot2new}):
\begin{equation}
\Psi_{N-n}(\vec{x}) = \sum\limits_{k = 0}^N\label{PsiNn}
b_{n,k}\Omega_k(\vec{x}).
\end{equation}
Formula (\ref{formal}) gives us the opportunity to express an
element $b_{m,p}$ by means of the $\tau^{(m)}$ matrices and an
arbitrary element $b_{m, N-m}$ on the crossed diagonal. This last
element can be fixed by the normalization condition for
$\Psi_{N-m}$. The reason for the ''inverted'' numeration of
$\Psi$ in (\ref{PsiNn}) is to make $\Psi_k$ dependent only on
$\Omega_l$; $l = 0,1,...,k$. In particular, $\Psi_0 \sim
\Omega_0$. So, applying the method of $SUSY-$separation of
variables to $\mathcal{H}^{(0)}$, $\widetilde{\mathcal{H}}^{(0)}$
one obtains a set of eigenvalues and eigenfunctions for
$\mathcal{H}^{(0)}$.

\par Keeping in mind that $s = B/\alpha > 0$, we can restrict
ourselves with $B>0,\quad \alpha>0$. The conditions of
normalizability of $\Omega_n$ (and therefore, of $\Psi_n$ ) for
all $\omega_n$ can be derived from the explicit expressions:
\begin{equation}
\Omega_n = \omega_n\exp{\kappa} = \left( \frac{(\cosh{(\alpha(x_1
+ x_2))} - 1)(\sinh{(\alpha(x_1 - x_2))})}{\sinh{(\alpha(x_1 + x_2
))(\cosh{(\alpha(x_1 - x_2))} - 1)}}
\right)^{A/(\sqrt{2}\alpha)}P^{s-n}_s(\xi_1)P^{s-n}_s(\xi_2).
\end{equation}
The constraint is: $ -\frac{1}{\sqrt{2}} < \frac{A}{\alpha} <
\frac{1}{\sqrt{2}}. $ The condition $A>0$ keeps the strength of
attractive singularities of both superpartners
$\mathcal{V}^{(0)}$, $\widetilde{\mathcal{V}}^{(0)}$ at $x_\pm
\rightarrow 0$ not exceeding the standard bound $-1/(4x_\pm^2)$.
The resulting range of parameters for both $\Omega_n$ and
$\widetilde{\Omega}_n$ is:
\begin{equation}
\alpha > 0;\qquad B>0;\qquad\frac{B}{\alpha} \in \mathbb{N};\qquad
0 < A < \frac{\alpha}{\sqrt{2}}.
\label{ranpar}
\end{equation}
\par At first it seems that the energy eigenvalues (\ref{eigenval}),
which were built above from the analysis of the zero-modes of
$\mathcal{Q}^+$, should be absent in the spectrum of its
superpartner $\widetilde{\mathcal{H}}^{(0)}$, since the corresponding
eigenfunctions are annihilated by $\mathcal{Q}^+$.
However, the whole procedure of $SUSY-$separation
of variables can also be implemented for the spectral problem for
$\widetilde{\mathcal{H}}^{(0)}$ by suitably replacing
(\ref{nom}) with $\mathcal{Q}^-\vec{\widetilde{\Omega}}$ = 0.
Since $\mathcal{Q}^-$ and $\mathcal{Q}^+$ differ only by sign in
front of the first derivatives (see (\ref{potphinew})), one
should use the ''gauge'' transformation with
$\exp{(-\kappa(\vec{x}))}$. In this case one will obtain the same
equations (\ref{eta}), as for the problem (\ref{nom}). Then the
zero-modes of $\mathcal{Q}^-$ can be written as:
$$
\widetilde{\Omega}_n(\vec{x}) =
\exp{(-\kappa(\vec{x}))}\omega_n(\vec{x}) =
\exp{(-2\kappa(\vec{x}))}\Omega_n(\vec{x}),
$$
and the corresponding matrix $\hat{C}$ is again triangular. To be
more precise, it is the same as (\ref{matc})-(\ref{matrc}) up to
the sign of the last term in (\ref{matc}). Therefore, its
eigenvalues, i.e. values of energy for
$\widetilde{\mathcal{H}}^{(0)}$, coincide with (\ref{eigenval}).
One can check that the eigenfunctions are normalizable in the
same range of parameters (\ref{ranpar}). Thus the obtained part
of the spectra of superpartners $\mathcal{H}^{(0)}$ and
$\widetilde{\mathcal{H}}^{(0)}$ {\bf totally} coincide. In a
certain sense this result is similar to one of the variants of the
second order intertwining in 1D HSUSY QM \cite{acdi}: the equal
number of bosonic and fermionic zero modes {\it does not} signal
the spontaneous breaking of the supersymmetry.

\subsection*{4.2. The method of shape-invariance}

\par Shape-invariance \cite{shinv}, \cite{cooper}, \cite{newmet}
is an additional property of intertwined superpartner Hamiltonians
which gives the opportunity to determine their spectra
algebraically. Namely, if both Hamiltonians depend on some extra
parameter (or set of parameters) $a$, this property reads:
\begin{equation}
\widetilde{H}(a_0) = H(a_1) + \mathcal{R}(a_0),\label{shinv}
\end{equation}
where $a_1 = f(a_0)$ is another value of parameter, and
$\mathcal{R}(a_0)$ does not depend on $\vec{x}$, i.e.
$\widetilde{H}$ has {\it the same} (up to an additive constant)
shape as $H$, but with another set of parameters.
\par Let us assume that we know some eigenfunction $\Psi^{(0)}$ of
$H$ and the corresponding eigenvalue $E^{(0)}$ in some range of
the parameter $a$. Starting from
\begin{equation}
H(a_1)\Psi^{(0)}(a_1) = E^{(0)}(a_1)\Psi^{(0)}(a_1)\label{2}
\end{equation}
and employing (\ref{shinv}), one obtains:
\begin{equation}
\widetilde{H}(a_0)\Psi^{(0)}(a_1) = (E^{(0)}(a_1) +
\mathcal{R}(a_0))\Psi^{(0)}(a_1).
\label{4}
\end{equation}
Using intertwining relations (\ref{intertw}) in (\ref{4}),
\begin{equation}
H(a_0)\left[ Q^-(a_0)\Psi^{(0)}(a_1)\right] = (E^{(0)}(a_1) +
\mathcal{R}(a_0))\left[ Q^-(a_0)\Psi^{(0)}(a_1)\right].
\end{equation}
This means that $H(a_0)$ has the eigenvalue $E^{(1)}(a_0) =
E^{(0)}(a_1) + \mathcal{R}(a_0)$ with the wave function
$\Psi^{(1)}(a_0) = Q^-(a_0)\Psi^{(0)}(a_1)$ (its normalizability is not
guaranteed). Starting from (\ref{2}) with parameter
$a_2 = f(f(a_0))$ and repeating the described procedure twice one can
find:
\begin{equation}
H(a_0)\left[ Q^-(a_0)Q^-(a_1)\Psi^{(0)}(a_2)\right] =
(E^{(0)}(a_2) + \mathcal{R}(a_1) + \mathcal{R}(a_0))\left[
Q^-(a_0)Q^-(a_1)\Psi^{(0)}(a_2)\right],
\end{equation}
which gives one more point $(\Psi^{(2)}(a_0), E^{(2)}(a_0))$ in the
spectrum of $H(a_0)$. The general formulas are
\begin{equation}
\Psi^{(n)}(a_0) =
Q^-(a_0)Q^-(a_{1})...Q^-(a_{n-1})\Psi^{(0)}(a_n), \label{psish}
\end{equation}
\begin{equation}
E^{(n)}(a_0) = E^{(0)}(a_n) + \sum\limits_{k =
0}^{n-1}\mathcal{R}(a_k). \label{esh}
\end{equation}
\par So, we have constructed a ''shape-invariance chain''
of eigenfunctions starting from one given. The natural idea is to
combine this method with $SUSY-$separation of variables (if the
Hamiltonians possess shape-invariance): having (N+1)
eigenfunction from $SUSY-$separation, we use each of them to
start the described above shape-invariance chain. This procedure
was implemented for the generalized 2D Morse potential
(\ref{morse}) in \cite{newmet}.
\par For the 2D-generalized P\"oschl-Teller potential the
situation becomes more complicate. Indeed, the
Hamiltonians $H^{(0)}$, $\widetilde{H}^{(0)}$ with potentials
(\ref{pot1}) are shape-invariant:
\begin{equation}
\widetilde{H}^{(0)}(\vec{x}; B, \alpha) = H^{(0)}(\vec{x}; B -
\alpha, \alpha) + 2\left( B^2 - (B - \alpha)^2)
\right),\label{shinvpt}
\end{equation}
where, in the notations introduced above, $a_0 \equiv B$; $a_1 =
f(a_0) \equiv B - \alpha$. But, contrary to the model
\cite{newmet}, the method of $SUSY-$separation of variables {\bf
does not} work here, since the zero modes $\omega_n$ are
unnormalizable for all values of the parameter $A$.
\par This obstacle can be overcome by exploring the relation
between systems (\ref{pot1}) and (\ref{pot2new}):
\begin{equation}
H^{(0)}(x_1, x_2) = \mathcal{H}^{(0)}(x_+, x_-),
\label{HH}
\end{equation}
where in the r.h.s. arguments $x_{1}, x_2$ are substituted by
$x_+, x_-$. Because a part of the spectrum (and the
eigenfunctions) of $\mathcal{H}^{(0)}$ was found by the method of
$SUSY-$separation of variables (Subsection 4.1.), one can use the
relation (\ref{HH}) to obtain the corresponding part of the
spectrum (and the eigenfunctions) for $H^{(0)}$. Then one can use
these data to start shape invariance chains for the system
$H^{(0)}$, $\widetilde{H}^{(0)}$ according to (\ref{psish}) with
operators $Q^-$. For example, starting from the first zero mode one obtains:
\begin{equation}
\Psi^{(n)}(x_1,x_2; B) =
Q^-(x_1,x_2; B)...Q^-(x_1,x_2; B-(n-1)\alpha)
\Omega_0(x_+,x_-; B-n\alpha). \label{unnorm}
\end{equation}

\par The general formula for the spectrum can be obtained from
(\ref{esh}), where the $E^{(0)}$'s for each chain are taken from
(\ref{eigenval}):
\begin{equation}
E_{mn} = 2\alpha^2\bigl[ m(2s-2n-m) + n(2s-n) \bigr] =
2\alpha^2(m+n)\bigl[ 2s - (m+n)\bigr],
\label{Emn}
\end{equation}
where $0<m<s$ corresponds to the number of the chain (number of
eigenfunction constructed by $SUSY-$separation), and $0<n<s$ in
order to keep positive all of $\mathcal{R}(a_k)$, $k = 0, ...
,(N-1)$, since the ground state energies $E^{(0)}(a_k) = 0$.
Comparing (\ref{Emn}) with (\ref{eigenval}) one will find that these
points of the spectrum coincide exactly ($k \equiv m+n$). But at
closer examination {\bf all} "wave functions" of the form
(\ref{unnorm}) with $n \geq 1$ are unnormalizable due to the singular behaviour
of the supercharges (\ref{pot1}) at $x_{1,2}\rightarrow 0.$
Thus the seeming $(k+1)-$fold degeneracy of $k-$th energy level in (\ref{Emn})
is spurious since only one of the solutions of Schr\"odinger equation
(namely, the linear combination of zero modes $\Omega_n$) is normalizable.
Therefore, in contrast to the method of $SUSY-$separation of variables,
the method of 2D shape invariance is {\bf powerless} to give {\bf normalizable}
shape invariance chains of wave functions for the 2D P\"oschl-Teller
potential.

\section*{5. Two-dimensional intertwining relations of more
than second order}

\par In this Section we will imply equivalence of
$H^{(0)}$ and $\mathcal{H}^{(0)}$ up to a change of variables for
the new construction. Due to (\ref{HH}), the intertwining
relation $\mathcal{H}^{(0)}\mathcal{Q}^- =
\mathcal{Q}^-\widetilde{\mathcal{H}}^{(0)}$ can be rewritten as:
$$
H^{(0)}\widetilde{\mathcal{Q}}^- =
\widetilde{\mathcal{Q}}^-\breve{\mathcal{H}}^{(0)},
$$
where $\mathcal{\widetilde{Q}}^\pm(x_1, x_2) =
\mathcal{Q}^\pm(x_+,x_-)$ and $\breve{\mathcal{H}}^{(0)}(x_1, x_2)
= \widetilde{\mathcal{H}}^{(0)}(x_+, x_-)$. Comparing it with
intertwining relations for the pair ($H^{(0)}$,
$\widetilde{H}^{(0)}$), one can conclude, that $H^{(0)}$ has two
{\it different} superpartners:
\begin{equation}
\breve{\mathcal{H}}^{(0)}
\stackrel{\widetilde{\mathcal{Q}}^\pm}{\leftarrow\rightarrow}
H^{(0)} \stackrel{Q^\mp}{\leftarrow\rightarrow}
\widetilde{H}^{(0)},\label{twosup}
\end{equation}
intertwined by {\it different} supercharges. Therefore, the
Hamiltonians $\breve{\mathcal{H}}^{(0)}$ and
$\widetilde{H}^{(0)}$ can be considered as superpartners
intertwined by the {\bf fourth order} operators
$\widetilde{\mathcal{Q}}^\pm Q^\mp$. This pair does not obey the
shape-invariance property.
\par Because, the Hamiltonian
$\widetilde{H}^{(0)}$ is shape-invariant (\ref{shinv}), one can
develop the construction (\ref{twosup}):
\begin{equation}
\breve{\mathcal{H}}^{(0)}(a_0)\stackrel{\widetilde{\mathcal{Q}}
^\pm(a_0)}
{\leftarrow\rightarrow}H^{(0)}(a_0)\stackrel{Q^\mp(a_0)}
{\leftarrow\rightarrow}\widetilde{H}^{(0)}(a_0)  = H^{(0)}(a_1) +
\mathcal{R}(a_0)\stackrel{\widetilde{\mathcal{Q}}^\mp(a_1)}
{\leftarrow\rightarrow}\breve{\mathcal{H}}^{(0)}(a_1) +
\mathcal{R}(a_0), \label{chain}
\end{equation}
where $a_0 = B,\quad a_1 = B - \alpha$ (see Subsection 4.2.). The
outermost operators in (\ref{chain}) are intertwined by {\bf
sixth order} supercharges according to:
\begin{equation}
\breve{\mathcal{H}}^{(0)}(a_0)\left[
\widetilde{\mathcal{Q}}^+(a_0)Q^-(a_0)
\widetilde{\mathcal{Q}}^-(a_1) \right] = \left[
\widetilde{\mathcal{Q}}^+(a_0)Q^-(a_0)
\widetilde{\mathcal{Q}}^-(a_1) \right]\left[
\breve{\mathcal{H}}^{(0)}(a_1) + \mathcal{R}(a_0) \right],
\end{equation}
and contrary to the previous, fourth order, case, obey the
shape-invariance property. Thus one can continue the construction
of the spectrum of $\breve {\mathcal{H}}^{(0)}$.

\appendix
\renewcommand{\theequation}{A\arabic{equation}}
\setcounter{equation}{0} \section*{Appendix. The general solution
of the functional equation}

\par Applying operator
$(\partial_1^2 - \partial_2^2)$ to both sides of (\ref{phimain}),
one has:
\begin{equation}
2\left (\frac{\tilde{\phi}'_1(x_1)}{\tilde{\phi}_1(x_1)}\partial_1
- \frac{{\phi}'_2(x_2)}{{\phi}_2(x_2)}\partial_2\right
)(\phi_-(x_-) - \phi_+(x_+))= \left
(\frac{{\phi}''_2(x_2)}{\phi_2(x_2)} -
\frac{{\phi}''_1(x_1)}{\phi_1(x_1)}\right )(\phi_-(x_-) -
\phi_+(x_+)),\label{PPhid}
\end{equation}
where the notation $\tilde{\phi}_1(x_1)=1/\phi_1(x_1)$ was used.
The general solution of (\ref{PPhid}) is:
\begin{eqnarray}
\phi_-(x_-) - \phi_+(x_+) = &
(\tilde{\phi}'_1(x_1)\phi'_2(x_2))^{-1/2} \Lambda\left (\int
\frac{\tilde{\phi}_1(x_1)}{\tilde{\phi}'_1(x_1)}dx_1+
\int\frac{\phi_2(x_2)}{{\phi}'_2(x_2)}dx_2\right ) \nonumber\\= &
\phi_1(x_1)(\phi'_1(x_1)\phi'_2(x_2))^{-1/2} \Lambda\left (\int
\frac{\phi_1(x_1)}{\phi'_1(x_1)}dx_1-
\int\frac{\phi_2(x_2)}{\phi'_2(x_2)}dx_2\right ),\label{PPhilphi-}
\end{eqnarray}
where $\Lambda$ is an arbitrary function. The corresponding
expression for $(\phi_-(x_-) + \phi_+(x_+))$ can be obtained
using initial Eq.(\ref{phimain}):
\begin{equation}
\phi_-(x_-) + \phi_+(x_+) =
\phi_2(x_2)(\phi'_1(x_1)\phi'_2(x_2))^{-1/2} \Lambda\left (\int
\frac{\phi_1(x_1)}{\phi'_1(x_1)}dx_1-
\int\frac{\phi_2(x_2)}{\phi'_2(x_2)}dx_2\right ).\label{PPhilphi+}
\end{equation}
Expressions for $\phi_{\pm}$ in terms of $\Lambda$ should
depend on proper argument, i.e. $\partial_\pm\phi_\mp = 0,$
leading to additional constraints for the function
$\Lambda$:
\begin{eqnarray} \frac{1}{2}\left (\frac{{\phi}''_2(x_2)}{\phi_2(x_2)
{\phi}'_2(x_2)} +\frac{{\phi}''_1(x_1)}
{\phi_1(x_1){\phi}'_1(x_1)}\right )\Lambda & = & \left
(\frac{1}{{\phi}'_1(x_1)} - \frac{1}{{\phi}'_2(x_2)}\right
)\Lambda ', \label{PPhisysl1}\\
\left ({\phi}'_1(x_1) + {\phi}'_2(x_2) -
\frac{\phi_1(x_1){\phi}''_1(x_1)} {2{\phi}'_1(x_1)} -
\frac{\phi_2(x_2){\phi}''_2(x_2)} {2{\phi}'_2(x_2)}\right
)\Lambda & = & \left (\frac{\phi_2^2(x_2)}{{\phi}'_2(x_2)} -
\frac{\phi_1^2(x_1)}{{\phi}'_1(x_1)}\right )\Lambda
'.\label{PPhisysl2}
\end{eqnarray}
Its trivial solution $\Lambda \equiv 0$ gives $\phi_+ = \phi_- = 0$,
$\phi_{1,2}$ - arbitrary, and the potentials (\ref{potphiold}) and
(\ref{potphinew}) are amenable to separation of variables.
\par Otherwise one can exclude $\Lambda$ from
(\ref{PPhisysl1})-(\ref{PPhisysl2}):
\begin{equation} \frac{{\phi}''_1(x_1)\phi_2^2(x_2)}{\phi_1(x_1)} -
\frac{{\phi}''_2(x_2)\phi_1^2(x_1)}{\phi_2(x_2)}=
2{\phi}'^2_2(x_2) - 2{\phi}'^2_1(x_1) +
\phi_1(x_1){\phi}''_1(x_1) - \phi_2(x_2){\phi}''_2(x_2).\label{12}
\end{equation}
Though there is no separation of variables in (\ref{12}), it will appear
after applying the operator $\partial_1\partial_2$, so that:
\begin{equation}
\frac{(\phi_1''/\phi_1)'}{(\phi_1^2)'}=
\frac{(\phi_2''/\phi_2)'}{(\phi_2^2)'}\equiv
2a=const.\label{13}
\end{equation}
Integrating, multiplying by $\phi'_{1,2}$, integrating again and
taking into account (\ref{12}), one obtains the general solution
of (\ref{phimain}) in the form:
\begin{equation}
{\phi}'^2_{1,2}=a\phi_{1,2}^4 + b\phi_{1,2}^2 + c;\qquad x =
\pm\int\frac{d\phi_1}{\sqrt{a\phi_1^4 + b\phi_1^2 + c}};\quad b,c=const.
\label{ell}
\end{equation}

\section*{Acknowledgements}
Authors are indebted to A.A. Andrianov for useful discussions,
and especially to D.N. Nishnianidze for careful reading of
manuscript and for derivation of the general solution of
Eq.(\ref{phimain}). P.V. is grateful to International Centre of
Fundamental Physics in Moscow and Non-profit Foundation "Dynasty"
for financial support. This work was partially supported by the
grant No.05-01-01090 of the Russian Foundation for Basic Research.

%\newpage

\end{document}